\documentclass[conference]{IEEEtran}

\usepackage[hidelinks]{hyperref}

\ifCLASSINFOpdf
  \usepackage[pdftex]{graphicx}
\else
\fi
\usepackage[cmex10]{amsmath}
\usepackage{url}

\begin{document}
%
\title{Visual Information Retrieval in Endoscopic Video Archives}

\author{\IEEEauthorblockN{Jennifer Roldan Carlos\IEEEauthorrefmark{1},
Mathias Lux\IEEEauthorrefmark{1},
Xavier Giro-i-Nieto\IEEEauthorrefmark{2}, 
Pia Munoz\IEEEauthorrefmark{1} and
Nektarios Anagnostopoulos\IEEEauthorrefmark{1}}
\IEEEauthorblockA{\IEEEauthorrefmark{1}Klagenfurt University\\
Klagenfurt, Austria\\ 
Emails: jroldancarl@gmail.com, mlux@itec.aau.at, piamunozt@gmail.com, nek.anag@gmail.com}
\IEEEauthorblockA{\IEEEauthorrefmark{2}Universitat Politecnica de Catalunya\\
Barcelona, Catalonia/Spain\\
Email: xavier.giro@upc.edu}}

\IEEEoverridecommandlockouts
\IEEEpubid{978-1-4673-6870-4/15/\$31.00~\copyright~2015 IEEE}

\maketitle

\begin{abstract}
In endoscopic procedures, surgeons work with live video strea\-ms from the inside of their subjects. A main source for documentation of procedures are still frames from the video, identified and taken during the surgery. However, with growing demands and technical means, the streams are saved to storage servers and the surgeons need to retrieve parts of the videos on demand. In this submission we present a demo application allowing for video retrieval based on visual features and late fusion, which allows surgeons to re-find shots taken during the procedure.
\end{abstract}


%
\IEEEpeerreviewmaketitle

\section{Introduction}
%
%

While maintaining large video archives is an expensive venture for clinics and hospitals, more and more countries require the storage of those videos for legal reasons. Therefore, a growth of video archives over the next years is expected, especially related to endoscopic videos. 
As a consequence clever methods for indexing and retrieval are needed. Users of such an archive should be able to retrieve information on specific procedures, types of procedures or similarities between different procedures with ad hoc searches. 

There are mostly two main approaches for the creation of stored endoscopic videos depending on the doctors in charge of the procedure. (i) Those surgeons who are aware of the space requirements of videos and the tedious work of identifying relevant section in hour long recordings, typically turn on and off recording to just document the most important steps or results of the procedure. (ii) Surgeons, who just want to document their procedures for legal reasons and are not bound to re-visit them later, record the whole procedures including even large parts of the preparations and clean-up afterwards, which are typically out-of-patient recordings of less importance. However, in both cases surgeons rely on the same \emph{photo function}, which allows them to grab a frame from the video stream and store it, ie. to put it in a report later on. 

\begin{figure}[t]
\centering
\includegraphics[width=\columnwidth]{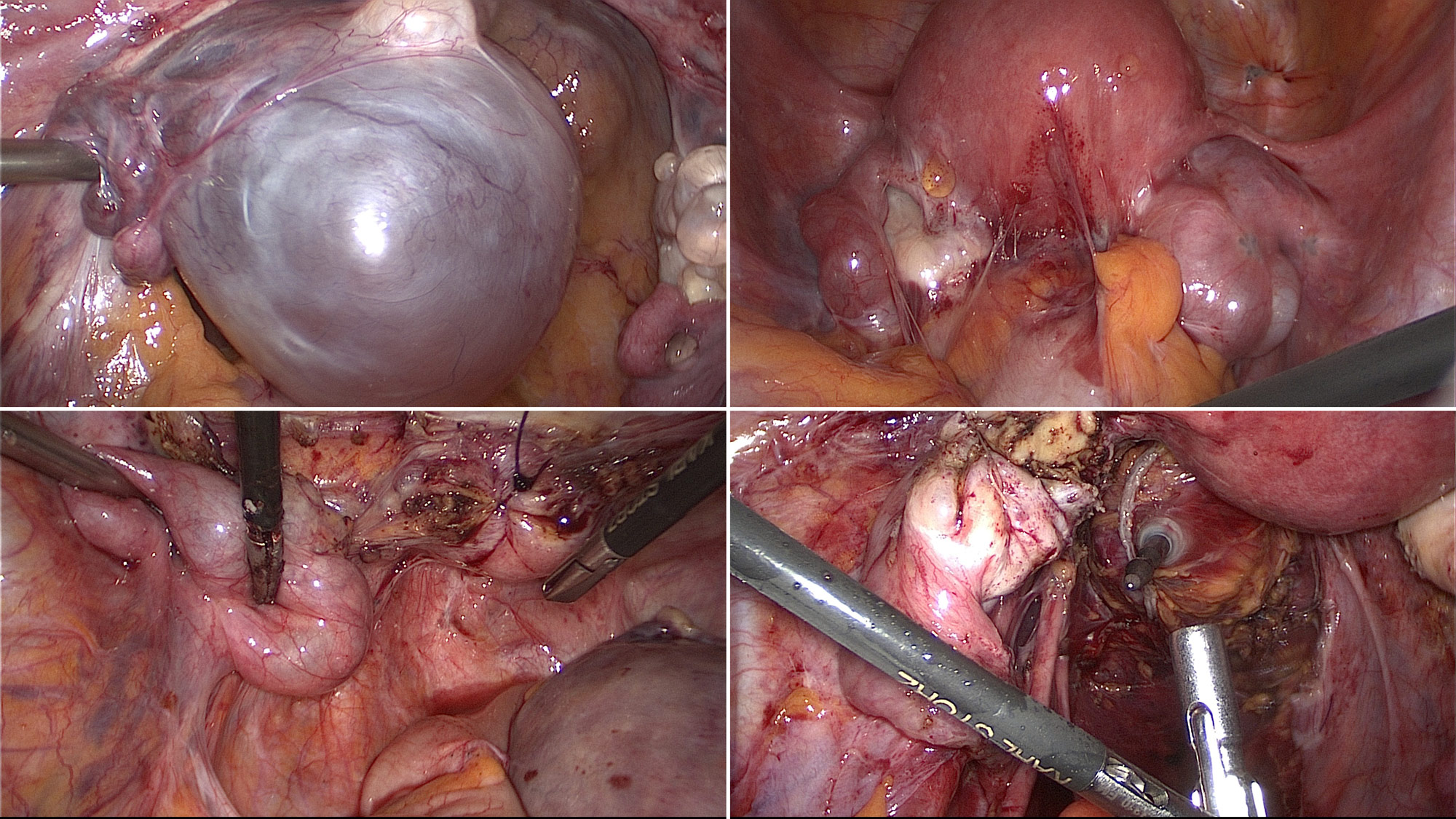}
\caption{Shots (photos) manually created from the surgeon in the course of the procedure.}
\label{fig:stills}
\end{figure}

In this paper we focus on the relation of \emph{photos} taken by a surgeon to the actual video streams as depicted in Fig.~\ref{fig:stills}. These photos, which we call \emph{shots} throughout the paper, are merely frames (still images) that have been saved at the time of operation on request of the surgeon, so they are also part of the video stream itself. Most important, what distinguishes them from the other frames of the video is that the surgeon intentionally directed the camera to a view to capture an optimal picture for later reference. 

In the framework presented in this paper, we focus on \emph{re-finding} those shots within video streams, i.e. we assume that the shots are known, but we want to (a) find the part of the original video where the shot was taken, and (b) find videos with visually similar frames to identify semantically similar scenes in different procedures. 
Ultimately, we believe such a system can be used for supporting medical research, education and training. We tested our application on a set of 1.276 videos ($\approx$ 33 hours) from 54 procedures.

The remainder of the paper is organized as follows. After surveying the most important related work, we first present the methods used for our approach, and then outline our application. After describing the test setup and presenting the results of a retrieval evaluation experiment and a qualitative result study, we conclude our paper and outline the next steps.


\IEEEpubidadjcol

\section{Related work}
The research field of image retrieval \cite{rui1999image, datta2008image} has extensively explored problems such as query by example or near-duplicate detection with high potential for the medical community.
In literature, a large number of research publications in medical imaging can be found, chiefly for gray scale images such as X-rays or magnetic resonance imaging (MRI).
\cite{n6}~describes potential applications of medical image retrieval and reviews some existing medical CBIR systems. \cite{n7}~also introduces different types of medical images used in CBIR systems as well as a large variety of techniques, potential applications and future lines. \cite{kumar2013content} provides a more recent review, emphasizing the multi-dimensional (2D and 3D) and multi-modality nature of the medical retrieval scenario.
Nevertheless, medical image and video retrieval remains an area of active research.

For example, the ImageCLEF benchmark \cite{imageclef} has created a strong community of researchers participating in the retrieval of medical images. A task for image-based retrieval was organized between 2004 and 2013. This case differs from the one addressed in this work because they were defined with 1-7 sample images accompanied by text. In the 2013 edition \cite{de2013overview}, the best textual run achieved the same performance as the best technique using both textual and visual features \cite{de2015comparing}. As in previous years, visual-only approaches achieved much lower results than the textual and multimodal techniques.
The best visual-based solution \cite{ozturkmenoglu2013demir} was based on the Color and Edge Directivity Descriptor (CEDD), a fuzzy color and texture histogram and a Color Layout Descriptor.

Content-based image retrieval in the medical domain has been addressed from low-level wavelet-based visual signatures \cite{quellec2010wavelet} to high level concept detectors \cite{rahman2011learning}. Another way to exploit visual features is to generate automatic text descriptors with computer vision algorithms \cite{kalpathy2010multimodal} and use these labels to support text-based queries. In \cite{schoeffmann2014keyframe} the authors focus on selecting frames from endoscopic video that describe the shot or frame sequence best. This work is relevant in the sense that while it does not offer a solution to the retrieval problem, it tries to  drastically reduces the visual information to the minimum number of necessary frames a surgeon needs for assessing the video.

Nowadays, medical retrieval systems have already become much more accessible on the web, typically supporting both textual and visual queries. These are the cases of NovaMedSearch~\cite{mourao2015multimodal} or GoldMiner~\cite{kahn2007goldminer}.

In contrast to most works on medical CBIR tasks, we address the problem of video retrieval, instead of still images. This venue has been previously explored in the literature. Specifically for real medical videos, \cite{n8}~proposes a framework that uses principal video shots for video content representation and feature extraction. The classification is mainly implemented by elementary semantic medical concepts, such as ``Traumatic surgery'' or ``Diagonosis''.
Moreover, \cite{n10} presents a framework to retrieve short videos in real time by modeling the motion content with a polynomial model. 





\section{Methods}
\label{sec:methods}

In our approach we focus on content based video indexing and retrieval to match example query content (still images) to target video content by extracting and indexing visual feature descriptors. For tests on the utility and usefulness of different approaches, we implemented three methods for visual retrieval: two of which use global features and feature fusion, and the third one which employs local features based on a recent model.

\subsection{Global and Local Features}

In our study we have tested three different types of global features: (i) \emph{Color and Edge Directivity Descriptor (CEDD)}~\cite{Chatzichristofis2008}, a compact joint histogram of fuzzy color and texture, (ii) the \emph{auto color correlogram}~\cite{Huang1997}, a color feature that measures how often a color encounters itself in a neighborhood, and (iii) the \emph{pyramid histogram of oriented gradients} (PHOG)~\cite{Bosch2007}, a fuzzy gradient histogram organized in a spatial pyramid.

A local feature solution has also been adopted to be compared with the global ones. We employ a localized version of CEDD using the SIMPLE model~\cite{Iakovidou2014} which has outperformed classical local features in many scenarios. SIMPLE uses a key point detector to find salient points on different scales. Based on the scale the point has been found, a local image patch is indexed with a compact and composite descriptor. Following that, the bag of visual words model is used to aggregate local features into histograms. For the experiments reported in this paper we used SIMPLE with the CEDD feature, the SURF key point detector~\cite{bay2006}, and k-means to create a visual vocabulary of 512 visual words, as a vocabulary of 512 visual words has been reported to lead to robust retrieval performance over several data sets in \cite{Iakovidou2014} for the SIMPLE approach.



\subsection{Late Fusion by Rank and by Score}

For fusion, each descriptor can be considered as an \emph{independent retrieval model}~\cite{Escalante2008}. To incorporate more characteristics than just one feature vector, independent retrieval models can be fused. Mainly, two types of fusion schemes are typically adopted. In \emph{early fusion} the different retrieval models and feature spaces are integrated from the start, and afterwards a multimodal representation is learned. \emph{Late fusion} approaches on the other hand infer similarity directly from unimodal features by creating a relevance score or ranked list for each of them, and integrate results at the end~\cite{Snoek2005} by fusing different scores or ranks. 

\begin{figure*}
	\includegraphics[width=\textwidth]{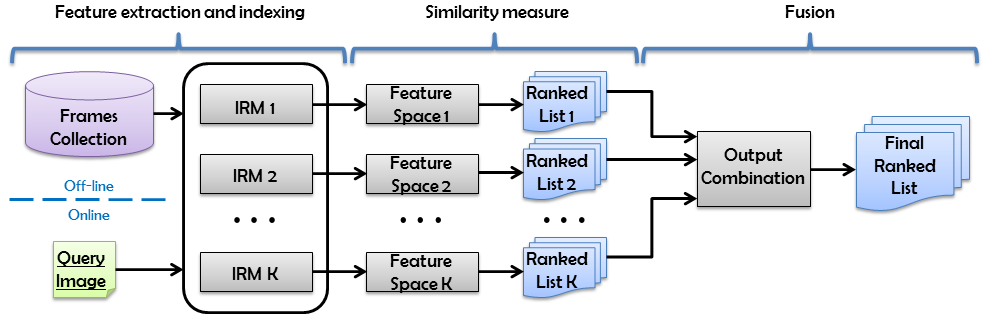}
	\caption{Application of late fusion in our approach, illustration is based on the work in \cite{Snoek2005, n3, n6}.}
	\label{fig:diagram}
\end{figure*}

Fig.~\ref{fig:diagram} shows the overall architecture. First, in an offline process, frames are collected and indexed. Based on the index and ad hoc search, similarity in different retrieval models is computed. 
A ranked list is generated based on each visual descriptor, and those lists are later fused in a single one.

In our approach, we employ a late fusion model based on multiple visual global features using a single query image. The objective of late fusion techniques is the combination and re-scoring or re-ranking of the initial result lists into one final list. Typically one truncates the initial lists to the top $N$ results and normalizes them either by rank  

$$\bar{R}_{k}(n)=\frac{N+1-R_{k}(n)}{N}  $$

or by score

$$\bar{R}_{k}(n)=\frac{R_{k}(n)-min(R_{k})}{max(R_{k})-min(R_{k})}$$

where $R_{k}$ is the initial result (rank or score) from the retrieval model $k$. For our approach we apply the sum approach, where either normalized ranks or normalized scores are summed up (cp. fusion strategies in~\cite{McDonald2005}), testing two approaches, sum of ranks and sum of scores: 

$$R_{t}(n) =\sum_{k} (R_{k}(n)) = R_{1}(n)+R_{2}(n)+...+R_{K}(n)$$

Note at this point that we did not investigate unimodal features as results in \cite{chatzichristofis2010latefusion} and \cite{Escalante2008} indicate that late fusion performs at least as good as the best unimodal feature of the ones that are fused.


\section{Our Application} 
\label{sec:application}


The goal of our application is to test and compare the different visual features and fusion methods presented in Section~\ref{sec:methods} for the retrieval of endoscopic videos. In particular, we addressed the use case of  re-finding shots within video streams with a query still image. 

This application was developed on a dataset of 1,276 video clips that were temporally sampled at 5 frames per second.
This dataset cannot be published due to confidentiality restriction given the medical nature of the data.
In order to define the experiments, we created a test dataset of query images. 
For this purpose, we used the shots generated by the surgeons during real procedures whenever they wanted to document a specific event that they consider important in the course of the surgery.
This way we exploited the interaction from experts in endoscopic videos to determine the highly informative frames in the video, assuming that given the original intention queries in a retrieval system would be from a similar nature. 
Notice that, as a result, our set of queries is a new group of images different from the uniformly sampled frames from the video dataset. Even more so, as the shots are taken from the live and not the recorded video, we assume that some of them are not even in the recorded clips. Using experts, we  cleaned out the query set aiming to remove stills that do not reflect a recorded video frame, ie. out-of-patient shots, survey shots, etc., resulting in 600 queries.

The test frames were indexed using the LIRE software library \cite{lux2013lire}, a highly versatile image retrieval engine that can extract and integrate up to 20 different visual features. All features and fusion strategies described in Section \ref{sec:methods} were implemented and assessed on this platform. Given that the reported experiments are a proof of concept, we did not explore at this stage additional indexing strategies such as index splitting, hashing or metric indexing.



\begin{figure*}[ht]
	\centering
		\includegraphics[width=\textwidth]{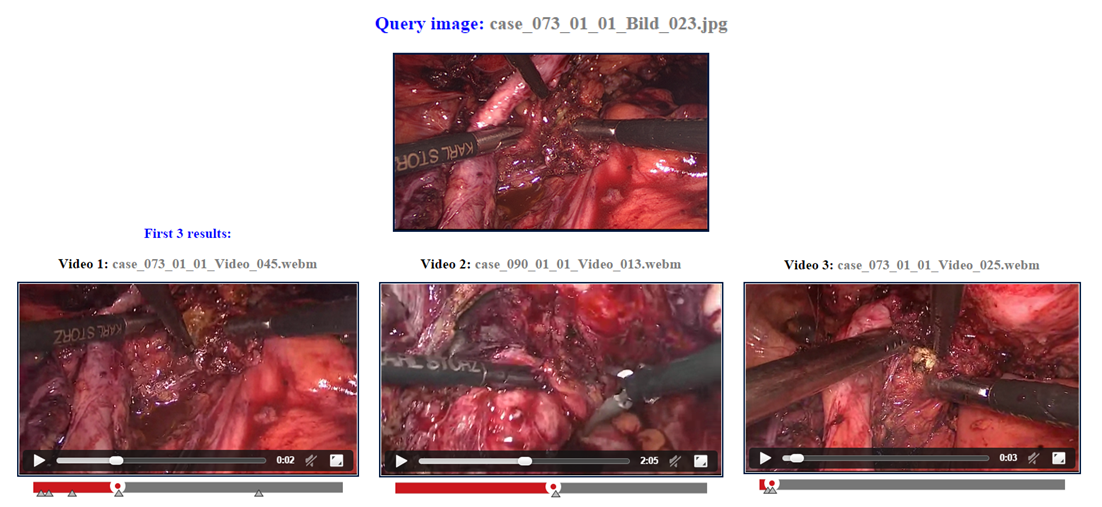}
	\caption{Screenshots of the result presentation showing the three top videos and the query image. All results are presented in HTML5 and can be viewed in recent browsers supporting HTML5 videos and JavaScript. Best matching frames are indicated by triangles in the red and gray time line below the video player.}
	\label{fig:screen}
\end{figure*}

Our application presents the results in a visual form in HTML5 for a recent version of common browsers. For each query an HTML file is generated displaying the query image, the list of similar images that the demo application finds, and the videos where both the query image and the rest of the frames belong to. All of the items appear along the time line where the images were taken. 
The screenshot presented in Fig.~\ref{fig:screen} shows the results of a shot query. Instead of showing the image results, only their positions in the video are indicated in the time line. Due to the nature of visual similarity search, retrieved frames look very much like the query, so showing them would not help the user in re-finding them in the video streams.

Based on the top 10 hits for each query, we determine the three best matching videos and present them to the user, highlighting the time location where the matching frames have been actually found, as shown in Fig.~\ref{fig:screen}. As the search process is based on frames within the videos and the result list is also composed of video frames, our system aggregates the frames as a last step. For this reason, the final ranked list of videos is based on their best matching frame, ie. the most similar frame defines the best matching video, the next most similar frame of a different video defines the second best matching video, etc.




\section{Evaluation}


Our data set covers roughly 33 hours of anonymized video data of laparoscopy procedures. For each of the procedures we had several shots manually taken by the surgeons. The videos were taken from different surgery’s cases of several patients. Due to the long duration of each intervention and the high resolution and bit rate of the videos, the whole surgery is divided in several videos, resulting in an overall file count of 1,276 videos. 
Due to the sheer size of the video archive, we employed temporal subsampling and extracted five frames per second for indexing, all in all 593,446 frames. Average linear search time for combining three retrieval models --  \emph{color and edge directivity descriptor} (CEDD), \emph{color correlogram}, and \emph{pyramid histogram of oriented gradients} (PHOG) -- was 30 seconds. The three retrieval models have been chosen as (i) CEDD is known to work in medical retrieval \cite{ozturkmenoglu2013demir}, (ii) color correlogram works extremely well compared to other global features \cite{lux2013lire} and (iii) PHOG covers similarity in detailed textures, which is not done by the other two features. Note that for this proof of concept we did not employ indexing strategies like hashing, metric indexes or clustering to speed up searching. 

For our experiments, we used 600 queries based on shots captured by the surgeons, as presented in Section~\ref{sec:application}.  Our experiments were twofold. First, we investigated the potential of each query to retrieve the video of the procedure where the query shot had been captured from. A quantitative metric was computed by comparing the retrieved videos with the ground truth. As our user interface only displays the top three ranked results, our study focused in the precision at positions 1, 2, and 3. 

As a second qualitative evaluation was ran with a \emph{thinking aloud test}~\cite{boren2000}. We created an interactive web page (cp. Fig.\ref{fig:screen}) featuring ten different surgery cases, and for each of them, the query shots available for search. The three search approaches were blindly labeled as search engine A (for sum of ranks fusion of global features), search engine B (for sum of scores fusion of global features) and search engine C (for the use of SIMPLE based local features). This way, we avoided any bias of the subjects towards any of the three approaches. 

\begin{figure}[ht]
	\centering
		\includegraphics[width=\columnwidth]{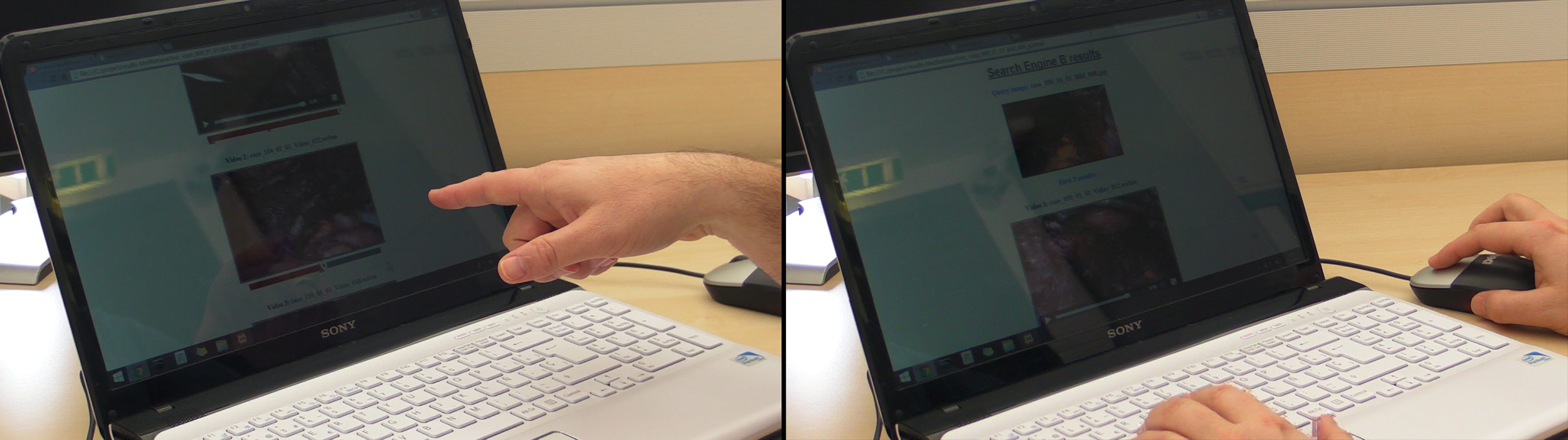}
	\caption{Still frames from the thinking aloud test recordings. Test participants pointed out and explained the utility of particular results.}
	\label{fig:test}
\end{figure}

We asked participants to investigate and compare the results of the different search engines and to give us feedback upon their quality and their usefulness. To allow participants to investigate subtle and non-obvious differences between the different search engines, we encouraged them to open multiple tabs in the web browser and compared the results by switching between them.  
We asked the users to test which of the three search engines satisfies the users’ needs, and which of them gives subjectively better results, ie. more accurate or broader. 
It was up to them to decide if the search engines returned what seemed natural to the users. It was up to the users to pick several of the queries and investigate possible results. In that sense it was a heuristic evaluation asking experts on the overall performance. 
The test subjects had been working in the field of computer science focusing on retrieval and analysis of endoscopic videos for several years. 
The participants were asked to voice their thoughts throughout the tests and the tests have been recorded on video (cp. Fig.~\ref{fig:test}). After the tests we reviewed and transcribed the interview recordings and test sessions. Based on the transcripts and the notes taken we discussed the results and concluded on the test.

\subsection{Experimental results}

Based on the whole set of queries, our tests have shown that for 470 out of 600 (78.3\%) of the queries, the source video was at the first position of the result list. In 84.2\% of the queries the source video was among the top three positions for the \emph{sum of ranks} approach, a very similar figure was obtained also for the \emph{sum of scores}. Local SIMPLE descriptor led to slightly better results, as in 79.8\% of the queries the source video was in the first place, while in 84.6\% of the queries the matching video was the first three videos (cp. Table~\ref{tab_test}).

\begin{table}[h]
\renewcommand{\arraystretch}{1.3}
\caption{Results of the tests on where that actual video can be found in the results. The first two columns give the two different tested feature fusion approaches, the third one gives the results on the use of the SIMPLE-CEDD descriptors.}
\label{tab_test}
\centering
\begin{tabular}{l|c|c|c|}
\cline{2-4}
                                      & \multicolumn{1}{l|}{\textbf{Sum of Ranks}} & \multicolumn{1}{l|}{\textbf{Sum of Scores}} & \multicolumn{1}{l|}{\textbf{SIMPLE-CEDD}} \\ \hline
\multicolumn{1}{|l|}{\textbf{Precision @ 1}} & 470                                      & 471                                         & 479                               \\ \hline
\multicolumn{1}{|l|}{\textbf{Precision @ 2}} & 21                                      & 20                                         & 21                               \\ \hline
\multicolumn{1}{|l|}{\textbf{Precision @ 3}} & 14                                      & 15                                         & 8                               \\ \hline
\end{tabular}

\end{table}

This indicates that the subsampling of five frames per second is enough for the used dataset to yield meaningful results. Note at that point that the shots are not necessarily in the video frames as they were taken from the live videos, so the ground truth at hand is more on a semantic level than mimicking a near duplicate task.

In the second experiment -- the thinking aloud test -- users in general expected to see the same background in several shots within the videos, which are similar to the query image. The participants choose the query image based on their intuition of what would result interesting, ie. they were driven by their own curiosity. They were driven by many reasons, as for example the simplicity of the background with specific organs on it, or specific movements of the surgeons as for instance cut tissue. Other reasons are a specific background, ie. bloody or damaged tissue, or a specific event using different instruments, which lets the user relate to a specific part of the procedure. Based on the overal state of tissue seen in the scene, ie. if it has been cut or cauterized, users know a rough time point within the surgery from the video. It gives them an orientation about the specific moment of the intervention, ie. they know whether the video is from in the beginning, during or the end of the procedure.
 After choosing a query image, the participants were expecting to see directly videos showing similar interventions. Due to the length of the videos, users consider a useful tool in the application when the results are marked in the time line; it allows them to find the right moment without the need to watch the whole video.

As an overall impression, for the search engines A and B, which are the sum of ranks and sum of scores fusion of global features, user commented they are good approaches showing in the top results the most relevant shots within the videos. However, in many cases the videos with higher ranks in the results show content which is semantically dissimilar by for instance featuring a different organ, instrument or background. For search engine C, which is based on the SIMPLE local features, users agreed it is the search engine that fits better when searching for semantically similar content. This technique also tends to retrieve fewer hits, which is (i) less confusing for the user and (ii) users need fewer steps to reach the right time point. 

As we indicated above, the dataset employed in this research is 33 hours approximately. 
Users considered search engine C a good approach because it only shows videos which contain real similarities with the query image, without showing false shots in the last positions. 
The participants indicate that this application is a good approach in order to re-find the video where the query image belong within the whole, eventually huge, data set. Mostly, this result appears in the first video of the list. They consider this a useful tool for the doctors, who day by day record a huge amount of data which is difficult to access and retrieve ad hoc when needed.

\section{Conclusion}

In this paper we presented a novel application for re-finding shots within endoscopic video streams, which is based on a real world use case from laporoscopic surgery. Two approaches novel for this domain have been tested. Late fusion of global features as well as the localized version of CEDD have not been applied to endoscopic video before. In our experiments we were able to find the shots in the respective videos within the first three results. A small study with two expert users also indicates that such a tool is of value for the everyday work routine of a surgeon. The methods employed, however, can be used in a number of scenarios. One obvious approach is video hyperlinking, ie. to find visually similar scenes in different video streams, and therefore, allowing for non-linear video browsing. Another interesting experiment would be to employ this approach to ad-hoc search within surgery procedures. Surgeons may take a shot and search the database for similar situations. 
Next steps in this project are a user study involving multiple surgeons, a large scale evaluation on our test data set including 600 shots. For deployment in real life, however, we have to investigate indexing strategies which allow for faster search time. We further aim at reducing the number of frames to be indexed by an automated method of frame selection for indexing, ie. by combining the work of \cite{schoeffmann2014keyframe} with ours.

\section*{Acknowledgements}
This work was supported by Lakeside Labs GmbH, Klagenfurt, Austria and funding from the European Regional Development Fund and the Carinthian Economic Promotion Fund (KWF) under grant KWF-20214/25557/37319. It has also been developed in the framework of the project TEC2013-43935-R, financed by the Spanish Ministerio de Economía y Competitividad and the European Regional Development Fund (ERDF).

\bibliographystyle{IEEEtran}
\bibliography{ref}

\end{document}